\documentclass[showpacs,aps,prd,nofootinbib,floatfix,amsmath,amssymb]{revtex4}
\usepackage{graphicx}
\usepackage[utf8x]{inputenc}
\usepackage{color}
\usepackage{subfig}
\usepackage{multirow}

\begin{document}

\title{\boldmath Prediction of $h\to\gamma Z$ from $h\to\gamma\gamma$ at LHC for the IMDS$_3$ Model}


\author{E. C. F. S. Fortes}
\email{elaine@ift.unesp.br}
\affiliation{
Instituto  de F\'\i sica Te\'orica--Universidade Estadual Paulista \\
R. Dr. Bento Teobaldo Ferraz 271, Barra Funda\\ S\~ao Paulo - SP, 01140-070,
Brazil
}

\author{A. C. B. Machado}%
\email{ana@ift.unesp.br}
\affiliation{
Instituto  de F\'\i sica Te\'orica--Universidade Estadual Paulista \\
R. Dr. Bento Teobaldo Ferraz 271, Barra Funda\\ S\~ao Paulo - SP, 01140-070,
Brazil
}

\author{J. Monta\~{n}o}
\email{montano@ift.unesp.br}
\affiliation{
Instituto  de F\'\i sica Te\'orica--Universidade Estadual Paulista \\
R. Dr. Bento Teobaldo Ferraz 271, Barra Funda\\ S\~ao Paulo - SP, 01140-070,
Brazil
}

\author{V. Pleitez}%
\email{vicente@ift.unesp.br}
\affiliation{
Instituto  de F\'\i sica Te\'orica--Universidade Estadual Paulista \\
R. Dr. Bento Teobaldo Ferraz 271, Barra Funda\\ S\~ao Paulo - SP, 01140-070,
Brazil
}

\date{09/25/15}
%


\begin{abstract}

We consider the decays $h\to\gamma\gamma,\gamma Z$ in the context of an extension of the standard model with two inert doublets and an additional $S_3$ symmetry. This model has contributions for these processes through new charged scalar-loops. Comparing our $h\to\gamma\gamma$ with the more precise available experimental data we can predict the behaviour of $h\to\gamma Z$ due that they depend on the same parameters, our estimation for this channel is 1.05 times the standard model value, but can be up to 1.16 if consider the $+1\sigma$ uncertainty from the $h\to\gamma\gamma$ data, and down to 0.96 if consider $-1\sigma$.

\end{abstract}


\maketitle

\section{Introduction}
\label{sec:intro}

The Large Hadron Collider (LHC) results indicate, for the first time, that at least one fundamental neutral scalar, here denoted by $h$, does exists in nature. Moreover, all its properties that have been measured until now are
compatible with the predictions of the standard model (SM) Higgs boson. For instance,  it is a spin-0 and
charge conjugation and parity symmetry (CP) even scalar~\cite{atlaspin,cmspin} and its couplings with gauge bosons and heavy fermions  are
compatible with those of the SM within the experimental error \cite{Aad:2012tfa,Chatrchyan:2012ufa}.
Notwithstanding, the data do not rule out the existence of new physics, in particular, processes induced at loop level have always been important to seek such evidence. This is the case of the decays $h\to\gamma\gamma$ and $h\to \gamma Z$ because they may have contributions from new charged particles. Recently, ATLAS and CMS Collaborations have measured the decay ratios for both processes~\cite{Aad:2014eha,Khachatryan:2014ira,Aad:2014fia,Chatrchyan:2013vaa}.
The decay of the Higgs boson into two photons is now in agreement with the SM prediction, if compared to 2012 data, but the decay into
a photon and a $Z$ has not been observed yet, however, ATLAS and CMS have presented upper limits for this decay, see Table~\ref{Atlas-CMS-Rates-TABLE}.

Moreover, motivated by physics of the dark matter (DM), neutrinos masses, hierarchy problem, and any
other physics beyond the SM, there are many phenomenological models that extend the scalar sector of the
SM with one or more scalar multiplets.
In fact, if in the future it becomes clear that dark matter consists of several components, multi-Higgs models will be  natural candidates. In particular $n$-doublet models with $n\geq2$, with or without scalar singlets and triplets,  will be interesting possibilities.  In particular,  the  inert Higgs doublet model (IDM) is the simplest model incorporating two DM candidates: one scalar and one pseudo-scalar field. A two inert doublet model can be obtained from a 3HDM plus a $Z_2$ symmetry with the inert doublets being odd and all the other fields are even under $Z_2$. Because of this symmetry, the two inert doublets do not get a vacuum expectation value (VEV), but the scalar potential is as complicated as the general 3HDM. The two inert doublets interact with each other as in the case of a general 2HDM i.e., 10 real dimensionless coupling constants, $\lambda$'s. Moreover each inert doublet interacts with the SM-like Higgs doublet as a 2HDM+$Z_2$ model, i.e., 10 $\lambda$'s more. It means that a two inert doublet model with just a $Z_2$ symmetry implies 23 real dimensionless parameters.  A more economical 3HDM with two doublets being inert can be built by imposing a $S_3$ symmetry. The $S_3$ symmetry allows that the symmetry eigenstates be related to the mass eigenstates through a tri-bimaximal-like matrix i.e., the mixing angles in all the scalar sectors are the same and of the Clebsch-Gordan coefficients type and there are no arbitrary mixing angles in the scalar sectors.
This is not the case in the general 3HDM with an arbitrary vacuum alignment. This sort of model was put forward for the first time in Ref.~\cite{Machado:2012ed} and the $h\to\gamma\gamma$ branching ratio in this context was considered in \cite{Cardenas:2012bg}. Here we will revisit this process with the more recent experimental data and also include the $h\to \gamma Z$ process.
We call this model IDM$S_3$ as in Ref.~\cite{Fortes:2014dca}, where we shown that the model has DM candidates. The Higgs mechanism provides a portal for communication between the inert sector and the known particles.

In the IDM and 3HDM$S_3$ the production of the 125 GeV Higgs is the same as in the SM, however the decays $h\to\gamma\gamma$ and $h\to \gamma Z$ can receive corrections due to the contributions of
charged scalars in loops. The phenomenology of IDM had been extensively discussed: i) in  the context of DM
phenomenology~\cite{LopezHonorez:2006gr,Hambye:2007vf,Dolle:2009ft,LopezHonorez:2010tb,Gustafsson:2012aj,Goudelis:2013uca}, ii) for collider phenomenology~\cite{Cao:2007rm,Lundstrom:2008ai,Swiezewska:2012eh} and, iii) IDM has been also advocated to
improve the naturalness idea~\cite{Barbieri:2006dq,Krawczyk:2013pea,Chen:2013vi}. However, all these references were published before the LHC data. The ratios of $h\to \gamma\gamma$ and $h\to \gamma Z$ were analyzed in the context of a general three Higgs doublet model in Ref.~\cite{Das:2014fea}. However these authors do not consider the case of two inert doublet and, unlike the present model, their model has arbitrary mixing matrices in the scalar sectors.

Special attention requires the $h\to\gamma Z$ rare decay since the current first attempt of measure this channel at LHC Run 1 shed an upper limit of one order of magnitude respect to the SM prediction ($R_{\gamma Z}$~=~1), see Table~\ref{Atlas-CMS-Rates-TABLE}. This is because the available luminosity at LHC is not sensitive enough to collect sufficient data of this process. Specifically, ATLAS~\cite{Aad:2014fia} has reported an upper limit of 11 times the SM expectation using a luminosity of 4.5 fb$^{-1}$ of $pp$ collisions at $\sqrt{s}=7$ TeV and 20.3 fb$^{-1}$ at
$\sqrt{s}=8$ TeV; CMS~\cite{Chatrchyan:2013vaa} reported an upper limit of 9.5 times the SM prediction, with integrated luminosities of 5.0 fb$^{-1}$ and 19.6 fb$^{-1}$ at $pp$ collisions of 7 TeV and 8 TeV, respectively. Nevertheless, the future of the detection of $h\to\gamma Z$ seems a difficult task according to the future LHC upgrades schedule \cite{Dawson:2013bba,LHC-Runs-Talks}: at LHC  Run 3 with 14 TeV will allow to collect 300 fb$^{-1}$ of data where the precision on the signal strength is expected to be $145-147\%$ at ATLAS and $54-57\%$ at CMS, and at Run 6 with 3000 fb$^{-1}$ the precision is expected to be $62\%$ at ATLAS and $20-24\%$ at CMS. Therefore, an accurate value for this decay will be one of the last data obtained by the LHC, but it is possible to predict the behavior of this decay from the process $h\to\gamma\gamma$ in the IDM$S_3$
due to the correlation of their common parameters,
specifically we estimate considering up to $\pm 1\sigma$ deviation from the experimental $R_{\gamma\gamma}$ data, that it is not possible a positive deviation larger than 1.16 times the SM value, nor a suppression beyond 0.96.

The outline of this paper is as follows. In Sec.~\ref{sec:model} we briefly present the model of Ref.~\cite{Machado:2012ed}.
In Sec.~\ref{sec:ratios} we calculate the  decays $h\to\gamma\gamma,\gamma Z$ in terms of the respective widths in the SM. The last section is designed for our conclusions. In the Appendix we present the amplitudes of the two processes and also details about the form factors and their solutions in terms of the Passarino-Veltman scalar functions and their analytical solutions.

\section{The Model}
\label{sec:model}

In \cite{Machado:2012ed} it was presented an extension of the electroweak standard model with three Higgs scalars, all of them transforming as doublets under
$SU(2)$ and having $Y=+1$. Some fields transform under $S_3$ as a doublet $D\equiv \textbf{2}$,  and some as a singlet $S\equiv\textbf{1}$. The scalar transform under $S_3$ as
\begin{eqnarray}
&&S=\frac{1}{\sqrt3}(H_1+H_2+H_3)\sim\textbf{1},\nonumber \\&&
D\equiv (D_1,D_2)=\left[\frac{1}{\sqrt6}(2H_1-H_2-H_3),\frac{1}{\sqrt2}(H_2-H_3)\right]\sim\textbf{2}.
\label{ma}
\end{eqnarray}
The vacuum alignment is given by $\langle H_1\rangle=\sqrt{3}v_{SM}$, and  $\langle H_2,H_3\rangle=0$ is an stable minimum of the potential at least at the tree level.

The most general scalar potential invariant under $SU(2) \otimes U(1)_Y \otimes S_3$ symmetry is given by:
\begin{eqnarray}
V(D,S) &=& \mu^2_sS^\dagger S+\mu^2_d [D^\dagger\otimes  D]_1 +\lambda_1
([D^\dagger\otimes  D]_1)^2
+  \lambda_2 [(D^\dagger\otimes D)_{1^\prime}(D^\dagger\otimes
D)_{1^\prime}]
\nonumber \\ &+&\lambda_3[(D^\dagger \otimes D)_{2^\prime}(D^\dagger\otimes D)_{2^\prime}]_1
+\lambda_4(S^\dagger S)^2+
\lambda_5[D^\dagger\otimes D]_1 S^\dagger  S+[\lambda_6 [[S ^\dagger D]_{2^\prime} [S^\dagger   D]_{2^\prime}]_1\nonumber \\ &+& H.c.]+
\lambda_7 S^\dagger [ D \otimes D^\dagger]_1 S +[\lambda_8[(S^\dagger\otimes D)_{2^\prime}(D^\dagger \otimes D)_{2^\prime}]_1+H.c.]
\label{potential1}
\end{eqnarray}
Denoting an arbitrary doublet by $\textbf{2}=(x_1,x_2)$, we have the product rule\textbf{S} as
$\textbf{2}\otimes\textbf{2}=\textbf{1}\oplus\textbf{1}^\prime\oplus
\textbf{2}^\prime$ where $\textbf{1}=x_1y_1+x_2y_2$,
$\textbf{1}^\prime=x_1y_2-x_2y_1$,
$\textbf{2}^\prime=(x_1y_2+x_2y_1,x_1y_1-x_2y_2$), and
$\textbf{1}^\prime\otimes\textbf{1}^\prime=\textbf{1}$~\cite{Ishimori:2010au}. Let us define $S=(s^+\,s^0)^T$, $D_i=(d^+_i\,d^0_i)^T,\;i=1,2$. In terms of the $S$ and $D_i$ fields, the potential in Eq.~(\ref{potential1}) is written as
\begin{eqnarray}
V(S,D_1,D_2) &=& \mu^2_sS^\dagger S+\mu^2_d (D^\dagger_1D_1+D^\dagger_2 D_2) +\lambda_1
(D^\dagger_1D_1+D^\dagger_2 D_2)^2
+  \lambda_2 (D^\dagger_1D_2-D^\dagger_2D_1)^2
\nonumber \\ &+&\lambda_3[(D^\dagger_1D_2+D^\dagger_2D_1)^2+(D^\dagger_1D_1-D^\dagger_2D_2)^2]
+\lambda_4(S^\dagger S)^2+
\lambda_5  (D^\dagger_1D_1+D^\dagger_2 D_2)S^\dagger  S \nonumber \\&+&[\lambda_6(S^\dagger D_1S^\dagger D_1+S^\dagger D_2S^\dagger D_2)+H.c.]
+
\lambda_7 S^\dagger (D_1D^\dagger _1+D_2 D^\dagger_2) S \nonumber \\&+& \lambda_8[S^\dagger D_1(D^\dagger_1D_2+D^\dagger_2 D_1)+S^\dagger D_2(D^\dagger_1D_1-D^\dagger_2D_2)+H.c.]
\label{potential2}
\end{eqnarray}
If $\mu^2_d>0$ only the singlet $S$ gain a VEV and if $\lambda_8=0$ this vacuum is stable at tree and the one-loop level. For this term be forbidden we impose a $Z_2$ symmetry under which $D\to -D$, and $S$ and all the other fields are even. The decomposition of the symmetry eigenstates we make as usual, as $H^0_i\!~=~\!(1/\sqrt{2})(v_i\!~+~\!\eta^0_i\!~+~\!i\,a^0_i),\;i=1,2,3$. We assume for the sake of simplicity that the VEVs are real and also iqual, i.em $v_1=v_2=v_3=v$.
these constraint equations  are reduced to a simple equation:
\begin{equation}
t_1 =t_2=t_3  =v(\mu^2_s  +    3\lambda_4 v^2),
\label{vinculos}
\end{equation}
and if $t_i=0$ we have $\mu^2_s= - 3 \lambda_4v^2 =- \lambda_4v_{SM}^2 <0$, which implies that $\lambda_4>0$.

the masses are given by:
\begin{eqnarray}\label{mrs1}
m^2_h= 2 \lambda_4v^2_{SM}, \qquad
m^2_{h_2}=m^2_{h_3}\equiv m^2_H=\mu^2_d+\frac{1}{2}\lambda^\prime v^2_{SM}, \\ \nonumber
m^2_{A_1}=0, \qquad
m^2_{A_2}=m^2_{A_3}\equiv m^2_A=\mu^2_d+\frac{1}{2} \lambda^{\prime \prime} v^2_{SM}.
\\ \nonumber
m^2_{h^+}=0,\qquad
m^2_{h_2^+}=m^2_{h_3^+}\equiv m^2_{h^+}=\frac{1}{4}(2\mu^2_d+\lambda_5v^2_{SM}).
\end{eqnarray}
Note that $\mu^2_d$ is not related to the spontaneous symmetry breaking and it is not protected by any symmetry, it may be larger than the electroweak scale.
As we see in Eq.~(\ref{mrs1}), $h^+$ and $A^0$ are the would-be Goldstone bosons that give masses to the $W^\pm$ and $Z$ gauge bosons and $h^0_{2,3}$ and $A^0_{2,3}$ are the inert fields.
Due to the S$_3$ symmetry and the vacuum alignment, we have a residual symmetry and due to it, the mass eigenstates of the inert doublets are degenerate, as we can see in Eq.~(\ref{mrs1}).

However, the residual symmetry, can be broken with soft terms in the scalar potential. So, adding the following quadratic terms $\nu^2_{nm}H^\dagger_nH_m$, $n,m=2,3$  and imposing that $\nu^2_{22}=\nu^2_{33}=-\nu^2_{23}\equiv \nu^2$,  the mass matrix will remain diagonalized by the matrix, so the inert character is maintained.
The eigenvalues are now:
\begin{eqnarray}\label{massfb}
&&\bar{m}^2_{h}=m^2_{h},\;\; \bar{m}^2_{h_2}=m^2_H ,\;\;
\bar{m}^2_{h_3}=m^2_H+ \nu^2,\nonumber \\&&
\bar{m}^2_{A_1}=0,\;\; \bar{m}^2_{A2}=m^2_A, \;\;\bar{m}^2_{A3}=
m^2_A+\nu^2,\nonumber \\&&
\bar{m}^2_{h_1^+}=0,\;\;\bar{m}^2_{h_2^+}=m^2_{h^+},\;\; \bar{m}^2_{h_3^+}=m^2_{h^+}+\nu^2,
\end{eqnarray}
where $m^2_h,m^2_A$, $m^2_{h^+}$ and $m^2_H$ are given in Eq.~(\ref{mrs1}).

The constraints from the vacuum stability as well as positivity on the relations of the couplings:
\begin{eqnarray}
&&\lambda_4  >  0,
\nonumber  \\  &&
\lambda_1 + \lambda_3  >  0 ,
\nonumber  \\  &&
\lambda_5   + 2 \sqrt{\lambda_4(\lambda_1 + \lambda_3)}  >  0,
\nonumber \\  &&
\lambda_5 + \lambda_7  - 2 \lambda_6 + 2 \sqrt{\lambda_4(\lambda_1 + \lambda_3)}  >  0 ,
\nonumber  \\ &&
\lambda_1 + \lambda_3  >  4 \lambda_2,
\label{nossa1}
\end{eqnarray}

In the lepton and quark sectors all fields transform as singlet under $S_3$, implying that they only  interact with the singlet $S$ as follows:
\begin{equation}
-\mathcal{L}_{Yukawa}=\bar{L}_{iL}(G^l_{ij}l_{jR}S+G^\nu_{ij}\nu_{jR}\tilde{S})+ \bar{Q}_{iL}(G^u_{ij}u_{jR}
\tilde{S}+G^d_{ij}d_{jR}S) +H.c.,
\label{yukawaint}
\end{equation}
$\tilde{S}=i\tau_2S^*$ and we have included right-handed neutrinos.
For more details see \cite{Machado:2012ed}.

The new inert scalar interactions with the gauge bosons, that arises from  $(D_{\mu} h_i)^\dagger (D^{\mu} h_i)$ with $i=2,3$, in the physical basis ($h_i=(h_i^+\ , \ h_i^0 )^T$) are given by
\begin{eqnarray}\label{intvh}
\mathcal{L}_{Gauge}&=& igs_W( \partial_\mu h_i^- h_i^+  -\partial_\mu h_i^+ h_i^-) A^\mu
    +igc_W \left( \frac{1 - t_W^2}{2} \right) (\partial_\mu h_i^- h_i^+  - \partial_\mu h_i^+ h_i^-) Z^\mu \nonumber\\
    && + i\frac{g}{\sqrt{2}} (\partial_\mu h_i^- h_i^0 - \partial_\mu h_i^0 h_i^-) W^{+\mu}
    -i \frac{g}{\sqrt{2}} (\partial_\mu h_i^+ h_i^0 -\partial_\mu h_i^0 h_i^+) W^{-\mu}
    \nonumber \\
    && + g^2 \, s^2_W \, h_i^- h_i^+ A_\mu A^\mu
    + g^2 c^2_W \left(\frac{1 - t_W^2}{2}\right)^2 h_i^- h_i^+ Z_\mu Z^\mu
    + 2 g^2\,  s_W c_W \left(\frac{1 - t_W^2}{2}\right) h_i^- h_i^+ A_\mu Z^\mu
    \nonumber \\
    && + \frac{g^2\,  s_W}{2} ( h_i^- W^+_\mu  +  h_i^+ W^-_\mu ) h_i^0 A^\mu
    + \frac{g^2c_W}{2} \left(\frac{1-t_W^2}{2}\right) (h_i^-W_\mu^+ + h_i^+W_\mu^-)h_i^0 Z^\mu.
\end{eqnarray}
The interactions between scalars in the physical basis are obtained from the following Lagrangian
\begin{eqnarray}\label{inthh}
\mathcal{L}_{Scalars} &=& -\lambda_4 v_{SM} h^3 -\frac{\lambda_5 v_{SM}}{2} (h_2^- h_2^+  + h_3^- h_3^+ )h
    -\frac{\lambda^\prime v_{SM}}{2} h  \left[(h_2^0)^2 + (h_3^0)^2\right]
    -\frac{\lambda_4 }{4} h^4
    \nonumber\\
    &&
    -\frac{\lambda^\prime}{2} h^2 \left[ (h_2^0)^2 +(h_3^0)^2\right] - 2 \lambda_3 h_2^0 h_3^0 h_2^- h_3^+ - (\lambda_1 + \lambda_3) (h_3^0)^2 h_3^- h_3^+
    \nonumber\\&&
    -(\lambda_2 + \lambda_3) (h_2^- h_3^+)^2  - (\lambda_1 + \lambda_3) (h_2^- h_2^+)^2
    -\frac{\lambda_1 + \lambda_3}{4} \left[ (h_2^0)^4 +(h_3^0)^4 \right]
    \nonumber\\
    && -(\lambda_1 + \lambda_3) (h_2^0)^2 (h_2^- h_2^+ + h_3^- h_3^+) \ ,
\end{eqnarray}
where in particular the terms proportional to $\lambda_5$ are the couplings between the SM-Higgs with the charged scalars involved in the $h\to\gamma\gamma , \gamma Z$ decays.

\section{Ratios $R_{\gamma\gamma}$ and $R_{\gamma Z}$}
\label{sec:ratios}

In this section we are going to study the ratios $R_{\gamma\gamma}$ and $R_{\gamma Z}$ predicted by the IDM$S_3$ respect to the SM.

To explore the sensitivity of the processes $h\to \gamma\gamma, \gamma Z$ due to new spin-0 content in the IDM$S_3$ we have used the experimental data reported by ATLAS and CMS collaborations. As can be seen in the
Table~\ref{Atlas-CMS-Rates-TABLE}, $h\to\gamma\gamma$ is within 1$\sigma$ related to the SM prediction, but for $h\to\gamma Z$ there is barely an upper limit of one order of magnitude above the SM prediction.
For the Higgs decay into two photons see the experimental Ref.~\cite{Aad:2014eha,Khachatryan:2014ira}, and for a photon and a $Z$ Ref.~\cite{Aad:2014fia,Chatrchyan:2013vaa}.

The $S_3$ symmetry and the vacuum alignment guarantee that the DM candidate does not decay into vector gauge bosons ($h_{2}^0\to \gamma\gamma, \gamma Z$) through quantum fluctuations induced by new charged spin-0 content, because it is forbidden the existence of the couplings $h_2^0h_2^+h_2^-$ and $h_2^0h_3^+h_3^-$, see Eq.~(\ref{inthh}), in contrast as it occurs with the SM-Higgs $h$ due to the presence of the couplings $hh_2^+h_2^-$ and $hh_3^+h_3^-$  that are proportional to $\lambda_5$.

As it is known, the Higgs discovery channel is $pp\to gg\to h\to\gamma\gamma$, and because of the nature of the IDM$S_3$ the SM interactions between the Higgs and quarks remain intact, thus there are no novelties in the Higgs fabric side $pp\to gg\to h$. On the other hand, new physics effects could come from new spin-0 particles in the Higgs decay process. More specifically, because the cross section for the Higgs production $pp\to gg\to h$ is the same for the SM and the IDM$S_3$, the application of the narrow width approximation (NWA) at the resonant point (when the gluon fusion energy is $\sqrt{\hat{s}}=m_{h}$), allow us to analyze the ratio signals with pure on-shell information
\begin{eqnarray}\label{Ratio}
R_{\gamma V} &\equiv& \frac{\sigma(pp\to gg\to h\to\gamma V)^{\textrm{IDMS}_3}}{\sigma(pp\to gg\to h\to\gamma V)^{\textrm{SM}}}
    \nonumber\\
    &\stackrel{\textrm{NWA}}{\simeq}& \frac{\sigma(gg\to h)^{\textrm{IDMS}_3}\textrm{Br}(h\to\gamma V)^{\textrm{IDMS}_3}}
    {{\sigma(gg\to h)^{\textrm{SM}}\textrm{Br}(h\to\gamma V)^\textrm{SM}}} \nonumber\\
    &=& \frac{\Gamma(h\to\gamma V)^{\textrm{IDMS}_3}}{\Gamma(h\to\gamma V)^\textrm{SM}}
    \frac{\Gamma_{h}^{\textrm{SM}}}{\Gamma_{h}^{\textrm{IDMS}_3}} \ ,
\end{eqnarray}
where $V\equiv \gamma, Z$.
We would like to call attention that in our scenarios the new neutral scalar masses forbid invisible decays of the SM-Higgs, except in the scenario 1a of Table 1 of Ref.~\cite{Fortes:2014dca} in which at the Born level yields $\Gamma(h\to h^0_3h^0_3)\sim 10^{-6}$ GeV, which is highly suppressed and does not disturb the total Higgs width, hence $\Gamma_{h}^{\textrm{IDMS}_3} \simeq \Gamma_{h}^{\textrm{SM}}$, leading to
\begin{equation}
\label{Ratio2}
R_{\gamma V} = \frac{\Gamma(h\to\gamma V)^{\textrm{IDMS}_3}}{\Gamma(h\to\gamma V)^\textrm{SM}} \ .
\end{equation}

As we have seen, the IDM$S_3$ gives rise to couplings between the new charged scalars and the SM-Higgs boson, and also with vector gauge bosons, but there are no modifications to the existing SM couplings, therefore for the decays $h\to\gamma\gamma, \gamma Z$ only a new scalar contribution is added to the existing ones.

\begin{figure}[!ht]
\subfloat[]{\includegraphics[width=4cm]{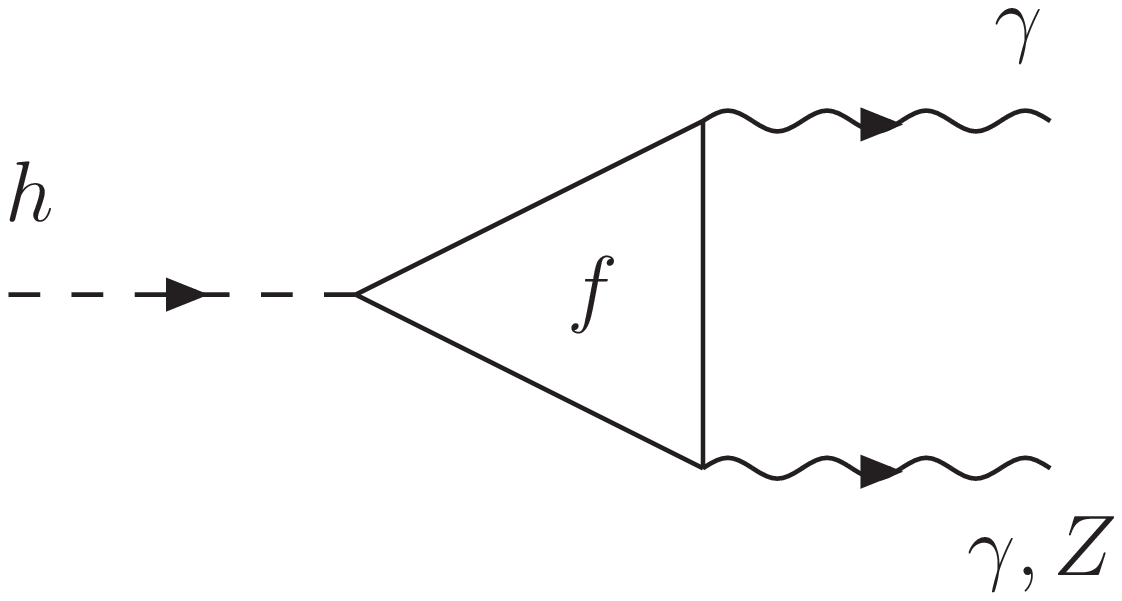}} \ \
\subfloat[]{\includegraphics[width=4cm]{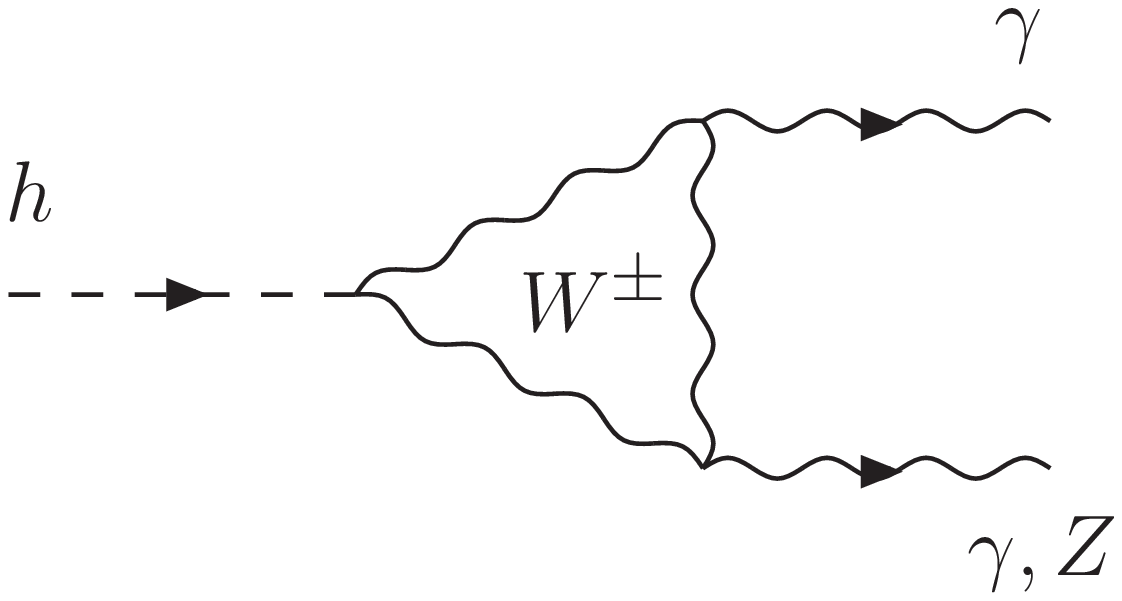}}  \ \
\subfloat[]{\includegraphics[width=4cm]{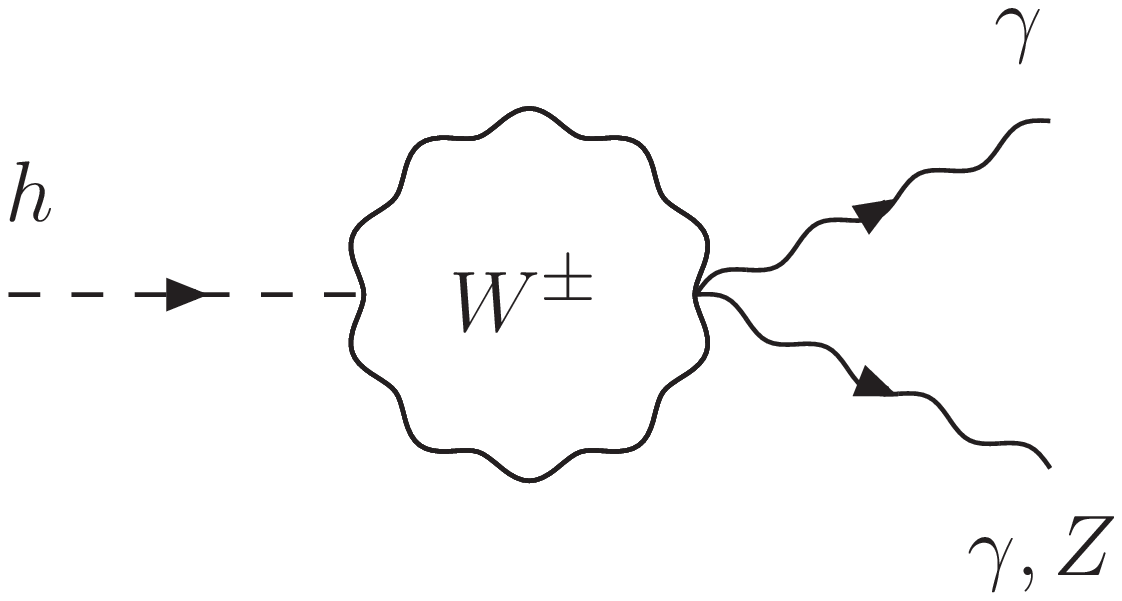}}  \\
\subfloat[]{\includegraphics[width=4cm]{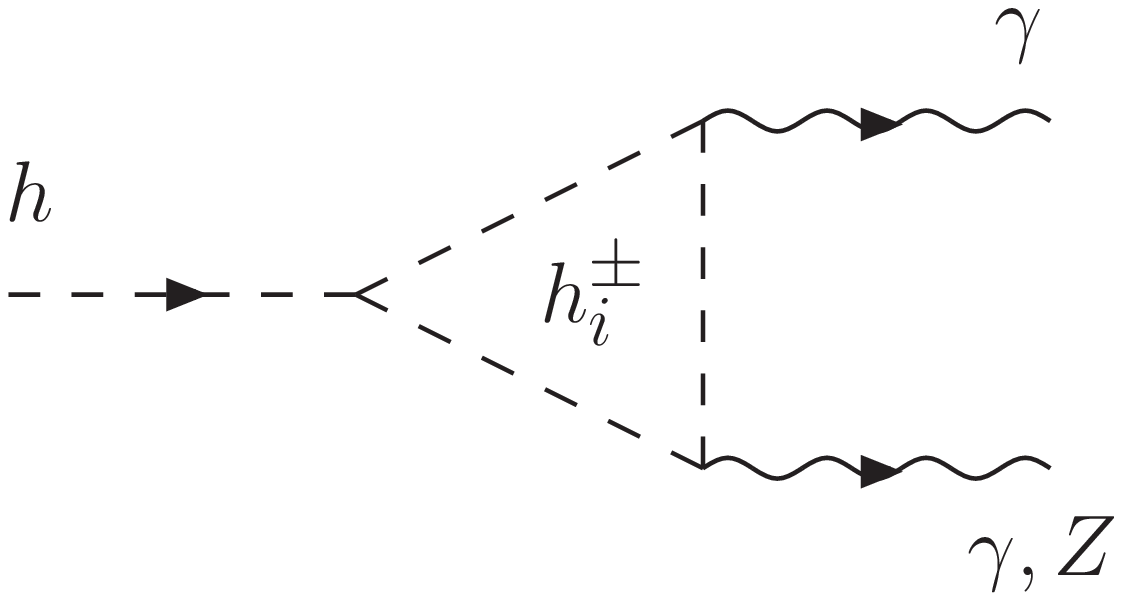}} \ \
\subfloat[]{\includegraphics[width=4cm]{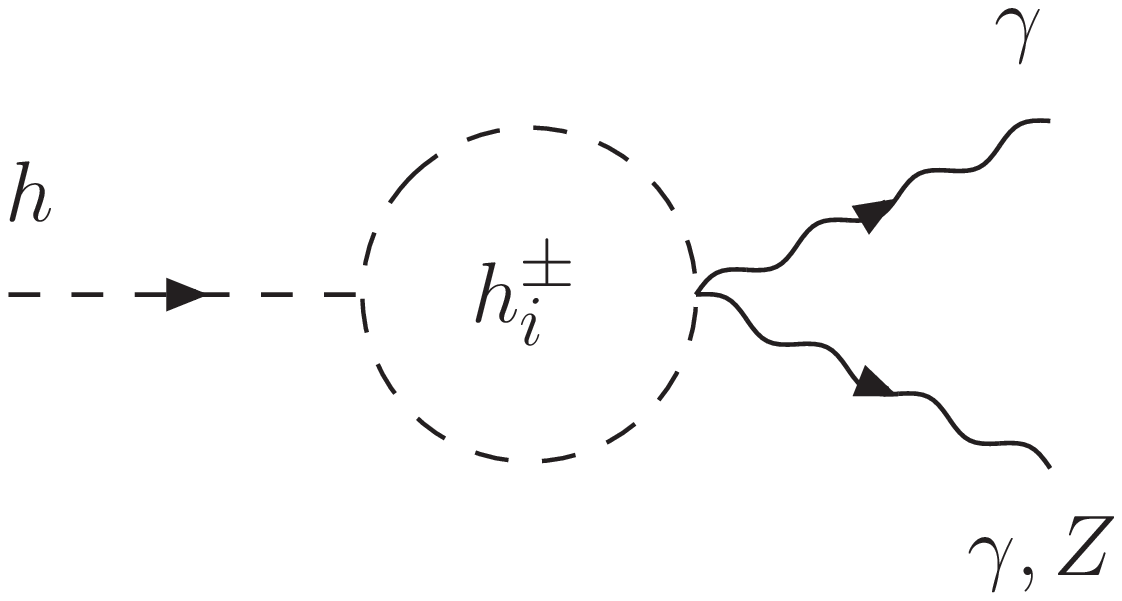}}
\caption{Decays $h\to\gamma\gamma, \gamma Z$.}\label{Decay-diagrams-h10-FIGURE}
\end{figure}

The participating diagrams in the processes $h\to \gamma\gamma, \gamma Z$ are illustrated in the Fig.~\ref{Decay-diagrams-h10-FIGURE} in the unitary gauge, where (a) corresponds to fermions, (b) and (c) to  $W$ gauge boson, and (d) and (e) to new charged scalars. We have constructed each diagram and performed the loop integrals with the Passarino-Veltman reduction method \cite{Passarino:1978jh} using the package \texttt{FeynCalc}~\cite{Mertig:1990an} which provides the results in terms of the scalar functions $B_0$ and $C_0$ \cite{'tHooft:1978xw}. We have  also calculated their corresponding general analytical solutions, which lead to the known standard notations of  Refs.~\cite{Gunion:1989we,Djouadi:2005gi,Djouadi:2005gj}.
Particularly here we work with the Djouadi notation \cite{Djouadi:2005gi,{Djouadi:2005gj}} for the width decays. In the Appendix we report the amplitudes of the processes and give details of the correspondence between our direct results in terms of the $B_0$ and $C_0$ functions and the Djouadi notation.

In the following we present the  decay widths showing explicitly only the new spin-0 contribution of the model. The other known spin-1/2 and spin-1 contributions are given in the Appendix.

The Higgs decay into two photons has new spin-0 contribution given by
\begin{equation}\label{width-h-2gamma}
\Gamma(h\to \gamma\gamma) = \frac{G_F\alpha^2m_{h}^3}{128\sqrt{2}\pi^3}
    \left| \sum_{i=1}^9 N_C^{f_i}Q_{f_i}^2 A_{1/2}^{\gamma\gamma}(\tau_{f_i}) +A_1^{\gamma\gamma}(\tau_W)
    +\frac{\lambda_5v_{SM}^2}{2}\sum_{i=2}^3\frac{1}{m_{h_i^+}^2}A_0^{\gamma\gamma}(\tau_{h_i^+}) \right|^2 \ ,
\end{equation}
with the form factors $A_{\textrm{Spin}}^{\gamma\gamma}$, where the charged scalar form factor is
\begin{equation}
A_0^{\gamma\gamma} \equiv -[\tau_{h^+}-f(\tau_{h^+})]\tau_{h^+}^{-2} \ .
\end{equation}
The $f(\tau)$ function is presented in the Appendix.

The Higgs decay into a photon and a $Z$ has also spin-0 contribution
\begin{eqnarray}\label{width-h-gammaZ}
\Gamma(h\to \gamma Z) &=& \frac{G_F^2m_W^2\alpha m_{h}^3}{64\pi^4} \left(1-\frac{m_Z^2}{m_h^2}\right)^3
    \left|\frac{2}{c_W} \sum_{i=1}^9 N_C^{f_i}Q_{f_i}g_V^{f_i} A_{1/2}^{\gamma Z}(\tau_{f_i}) +A_1^{\gamma Z}(\tau_W) \right. \nonumber\\
    && \left. +\frac{\lambda_5v_{SM}^2v_{h^\pm}}{2}\sum_{i=2}^3\frac{1}{m_{h_i^+}^2}A_0^{\gamma Z}(\tau_{h_i^+}) \right|^2 \ ,
\end{eqnarray}
where $v_{h^{\pm}} \equiv c_W(1-t_W^2)$, $A_{\textrm{Spin}}^{\gamma Z}$ are the form factors, with the new charged scalar contribution
\begin{equation}
A_0^{\gamma Z} \equiv -I_1 \ .
\end{equation}
See the Appendix for detailed information about all the form factors, the $I_{1,2}$ auxiliary definitions and also the $f(\tau)$ and $g(\tau)$ functions and their relations with the Passarino-Veltman scalar functions.

In the next section we report our phenomenological analysis for $h\to \gamma\gamma, \gamma Z$.
We use the values $m_h$= 125.09 GeV with the more recent data from PDG Live:
$m_W$= 80.385,
$m_Z$= 91.1876,
$m_u$= 0.0023,
$m_d$= 0.0048,
$m_s$= 0.095,
$m_c$= 1.275,
$m_b$= 4.18,
$m_t$= 173.07,
$m_e$= 0.000511,
$m_\mu$ = 0.105658,
$m_\tau$ = 1.77682,
all values in GeV,
$G_F$= 1.1663787$\times10^{-5}$ GeV$^{-2}$.

\begin{figure}[!ht]
\includegraphics[width=8.5cm]{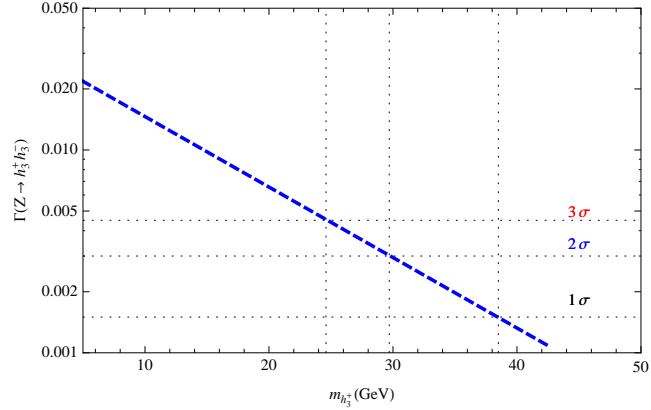}
\caption{$Z$ invisible decay width, as a function of the charged scalar $h^+_3$ mass. Imposing the error of the current value for the invisible decay as the allowed limit for the decay width, we a obtain a lower limit for the charged mass of 25 GeV.}\label{FIGURES-largurainvisivel}
\end{figure}

The four collaborations of LEP \cite{Abbiendi:2013hk} and ATLAS \cite{Aad:2013hla} have searched for charged scalars, notwithstanding, their lower limits depend on the model which is always the two Higgs doublet model (2HDM).  In LEP experiments, the searches include 2HDM of type I and II. Type I is searched in the ATLAS experiment. Both searches depend on the assumed branching ratio of the charged Higgs boson decays. ATLAS, for instance, assume $H^+\to c\bar{s}=100$\%. Summarizing, ATLAS has observed no signal for $H^+$ masses between 90 GeV and 150 GeV, and LEP has excluded this sort of scalars with mass below 72.5 GeV for  type I scenario and 80 GeV for the type II scenario. However, none of these results apply to our model since the charged scalar are inert and do not couple to fermions. Anyway, we will use 80 GeV for the mass of $h^+_2$ which is in the range of LEP and ATLAS results. For the other charged scalar, $h^+_3$ we will obtain a lower limit for its mass using its contribution to the $Z$ boson invisible decay width, where we have found $m_{h_3^+}> 25$ GeV, if we consider a 3$\sigma$ deviation for the invisible decay width  in our calculations. These results can be appreciated in Fig.~\ref{FIGURES-largurainvisivel}.

We first report the $h\to \gamma\gamma$ channel, and for the experimental comparison we use the data provided by the ATLAS \cite{Aad:2014eha} and CMS \cite{Khachatryan:2014ira} collaborations, given in Table~\ref{Atlas-CMS-Rates-TABLE}. Specifically, we follow the more stringent data which is reported by CMS, we explore its deviations values until $\pm 3\sigma$.

\begin{figure}[!h]
\subfloat[]{\includegraphics[width=7.5cm]{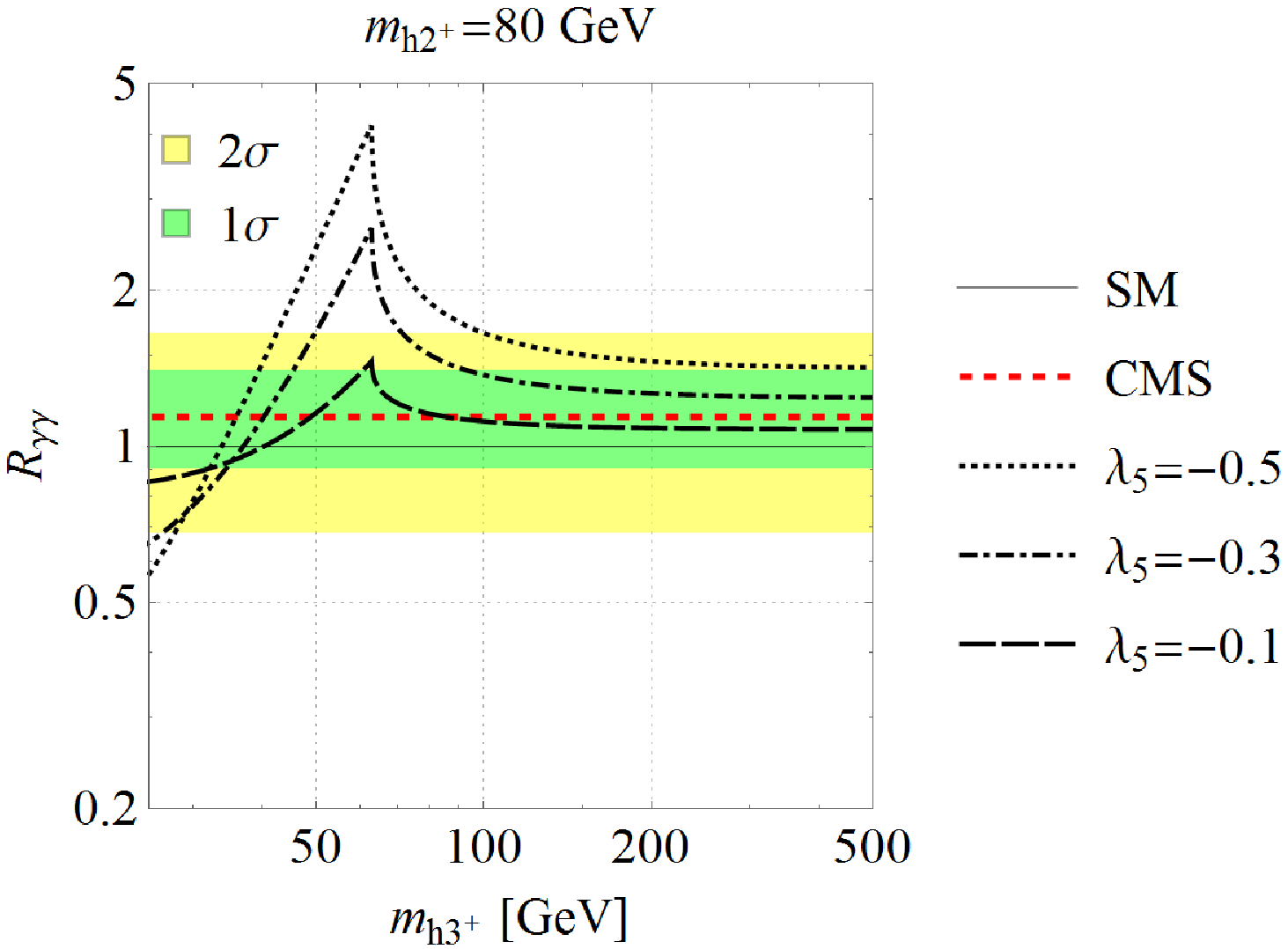}}
\subfloat[]{\includegraphics[width=7.5cm]{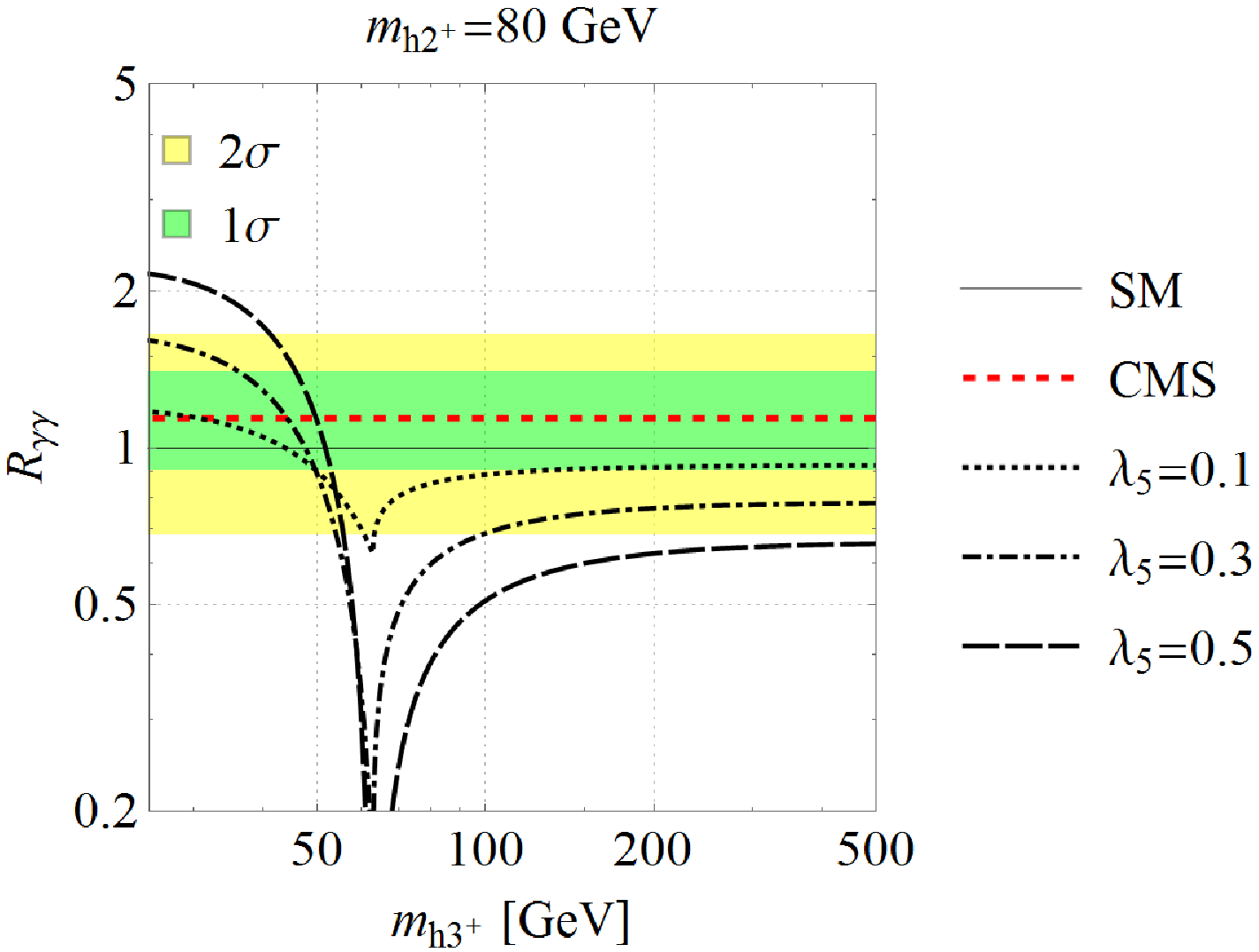}} \\
\subfloat[]{\includegraphics[width=8.5cm]{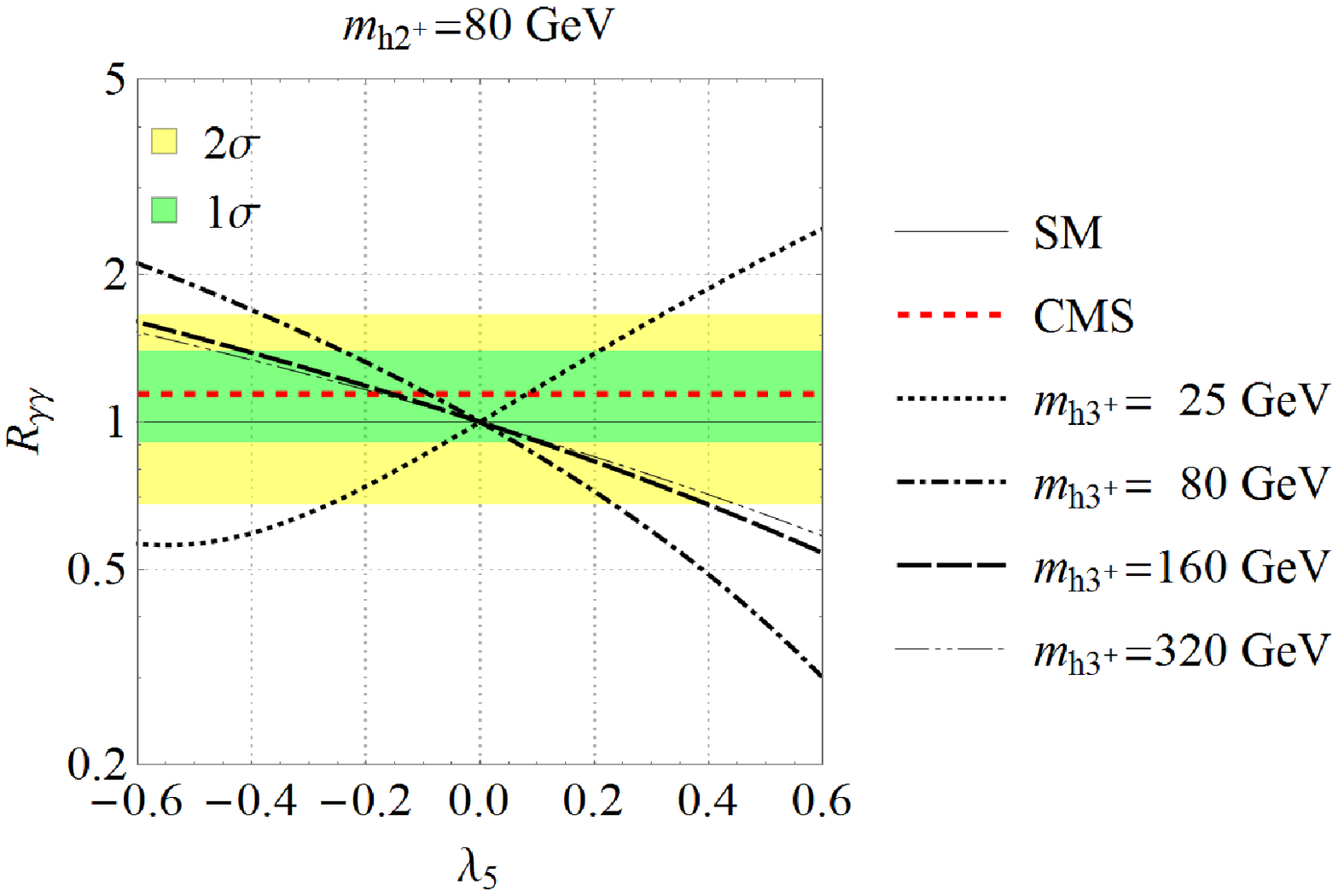}}
\caption{$R_{\gamma\gamma}$ with $m_{h_2^+}=$ 80 GeV. In (a) and (b) $R_{\gamma\gamma}$ as function of $m_{h_3^+}\geq 25$ GeV, but in (a) $\lambda_5$ is negative and in (b) is positive. In (c) $R_{\gamma\gamma}$ is presented as function of $-0.6\leq\lambda_5\leq 0.6$ with different values of $m_{h_3^+}$. }
\label{FIGURES-Results-Set-1}
\end{figure}

In the Fig.~\ref{FIGURES-Results-Set-1} we present $R_{\gamma\gamma}$ with $m_{h_2^+}=$ 80 GeV. First, we show $R_{\gamma\gamma}$ as function of $m_{h_3^+}$, in Fig.~\ref{FIGURES-Results-Set-1}(a) we consider $\lambda_5$ negative and in Fig.~\ref{FIGURES-Results-Set-1}(b) positive;
in Fig.~\ref{FIGURES-Results-Set-1}(c) $R_{\gamma\gamma}$ is presented as function of $-0.6\leq\lambda_5\leq 0.6$ and different values of $m_{h_3^+}$ are chosen.
From the three plots it can be appreciated that negative values of $\lambda_5$ and $m_{h_3^+}>m_h/2$ favors a positive deviation, being more compatible with the experimental allowed region if $m_{h_3^+}>$ 80 GeV.
For positive values of $\lambda_5$ and $m_{h_3^+}<m_h/2$ there is also a compatible positive deviation, but this mass scenario for the charged scalars could not be valid if the experimental values for one charged scalar mass limit from LEP \cite{Abbiendi:2013hk} and ATLAS \cite{Aad:2013hla} are also valid for an extra charged scalar $h_3^+$. If future
experimental data confirms a small negative deviation for $R_{\gamma\gamma}$, the $S_3$ model still has room for consistency with a scenario of positive $\lambda_5$ and $m_{h_{2,3}^+}>$ 80 GeV.

\begin{figure}[!h]
\begin{center}
\subfloat[]{\includegraphics[width=8.5cm]{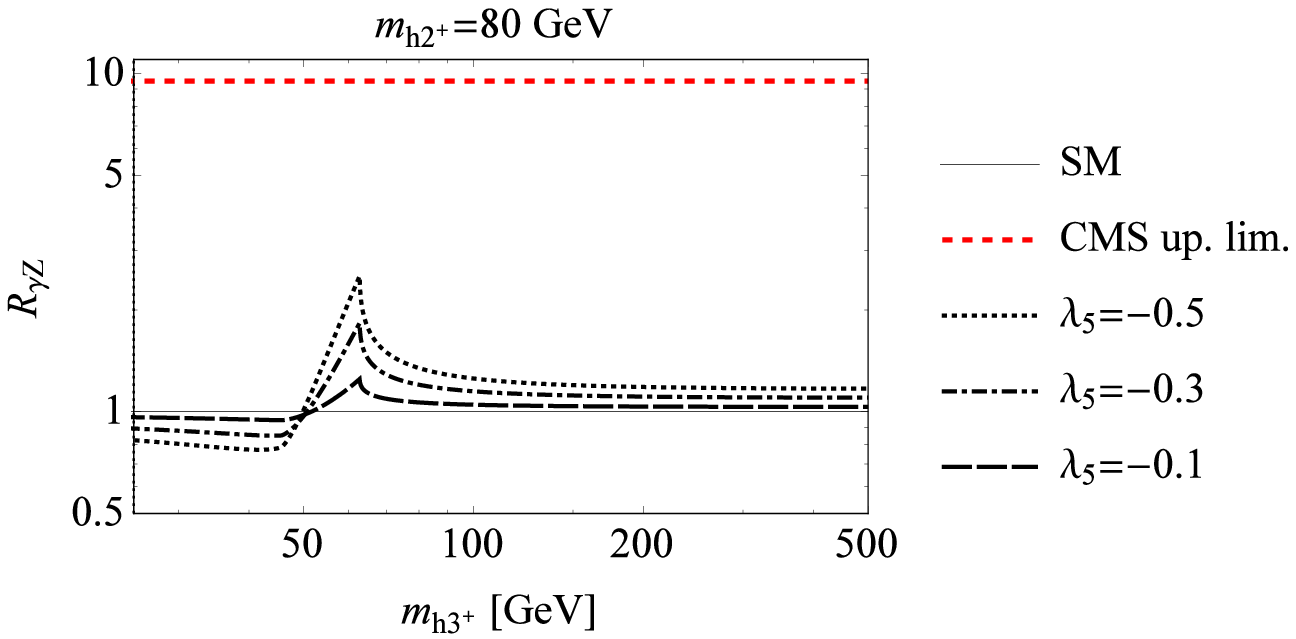}}
\subfloat[]{\includegraphics[width=8.5cm]{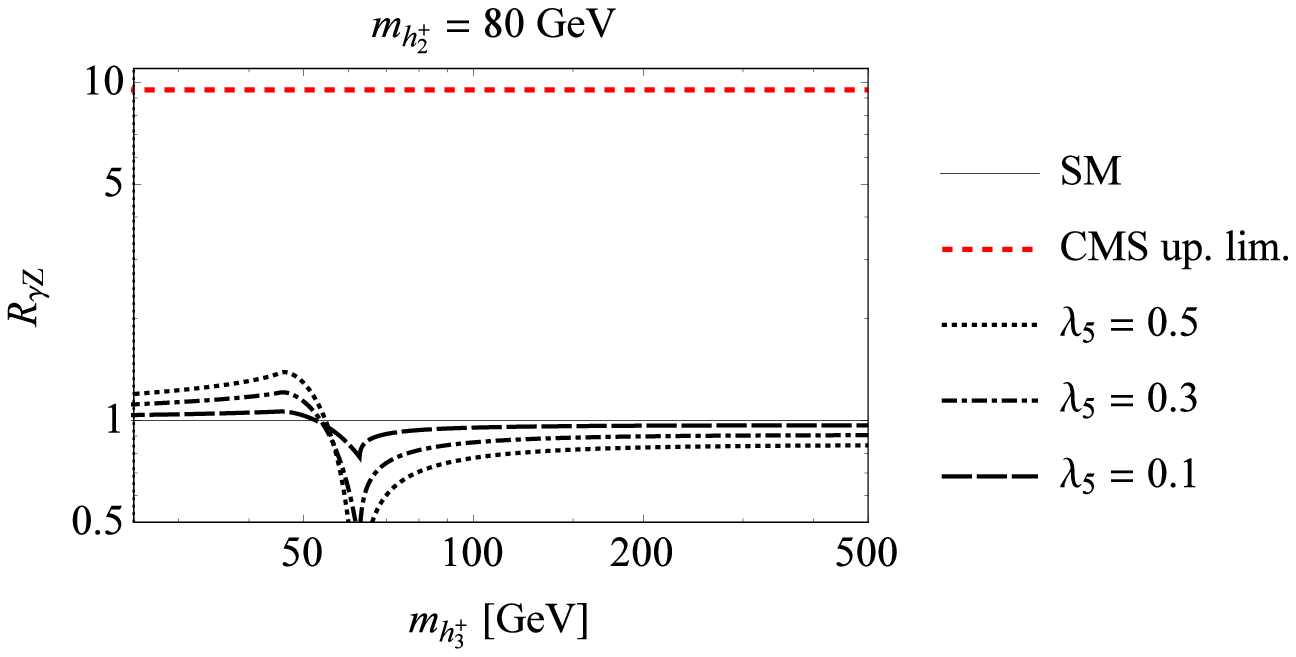}} \\
\subfloat[]{\includegraphics[width=8.5cm]{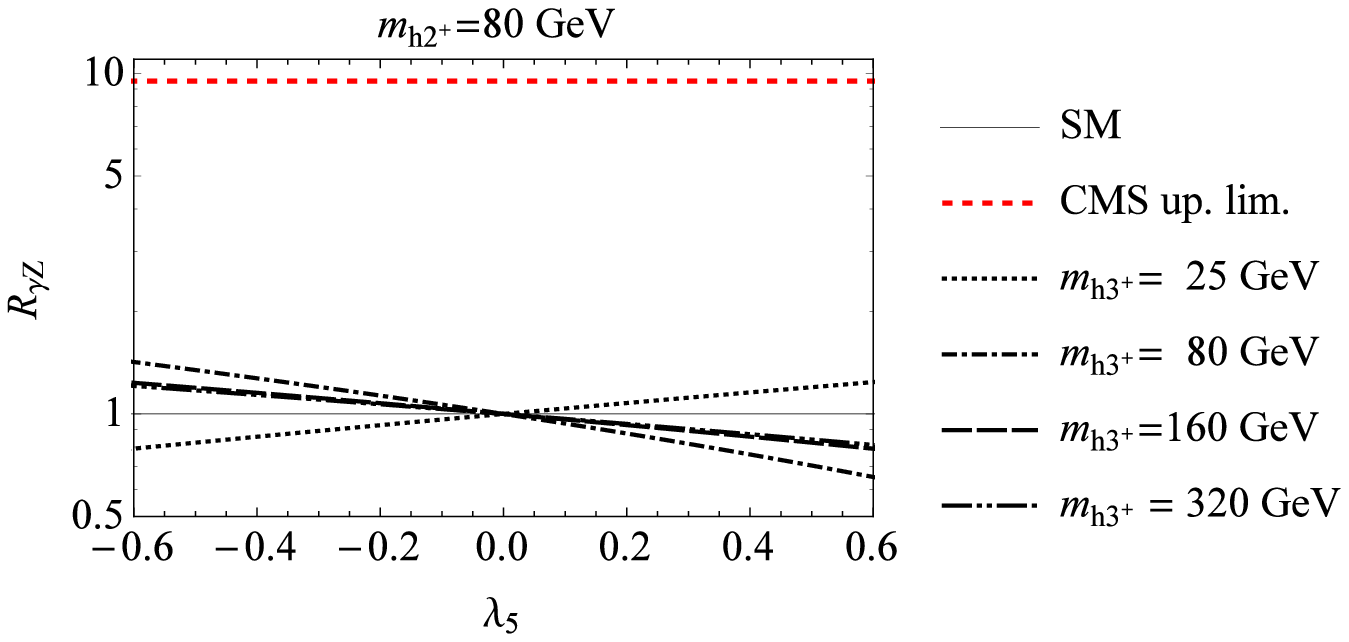}}
\caption{$R_{\gamma Z}$ with $m_{h_2^+}=$ 80 GeV. In (a) and (b) $R_{\gamma Z}$ as function of $m_{h_3^+}\geq 25$ GeV, but in (a) $\lambda_5$ is negative and in (b) is positive. In (c) $R_{\gamma\gamma}$ is presented as function of $-0.6\leq\lambda_5\leq 0.6$ with different values of $m_{h_3^+}$.}\label{FIGURES-Results-Set-2}
\end{center}
\end{figure}

Considering now $R_{\gamma Z}$, we have also made an analysis entirely analogous to the two photons case. The available experimental data for the process $h\to\gamma Z$ is still very rough, the ATLAS~\cite{Aad:2014fia} and CMS~\cite{Chatrchyan:2013vaa} reports provide so far upper limits of one order of magnitude larger than the SM prediction, see Table~\ref{Atlas-CMS-Rates-TABLE}. In Fig.~\ref{FIGURES-Results-Set-2} we illustrate the $R_{\gamma Z}$ results, this decay has almost the same shape and behavior than the two photons channel, except that now the signal is more suppressed considering the same parameters $\lambda_5$ and $m_{h_{2,3}^+}$. This result is congruent because $h\to\gamma\gamma$ has massless particles in the final state while $h\to\gamma Z$ produces one heavy particle, therefore it is expected that the latter process be less sensitive to the common parameters. Therefore, in our results, all analysis applied to $h\to \gamma\gamma$ also apply analogously to $h\to \gamma Z$, where the scenario of negative $\lambda_5$ and $m_{h_{2,3}^+}>$ 80 GeV agrees mostly with the more accurate experimental data for the two photons channel.

\begin{figure}[!h]
\begin{center}
\subfloat[]{\includegraphics[width=8cm]{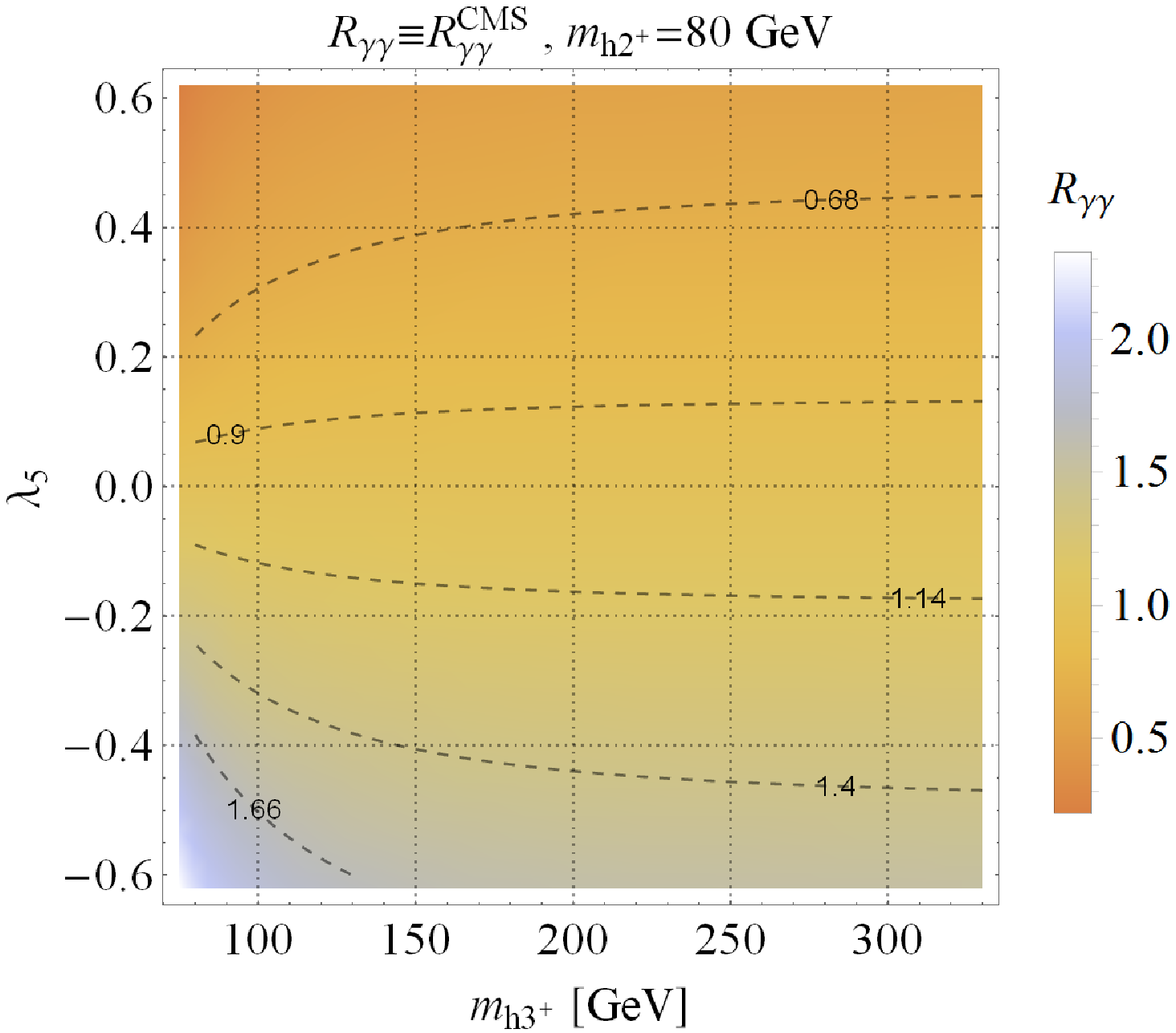}}
\subfloat[]{\includegraphics[width=8cm]{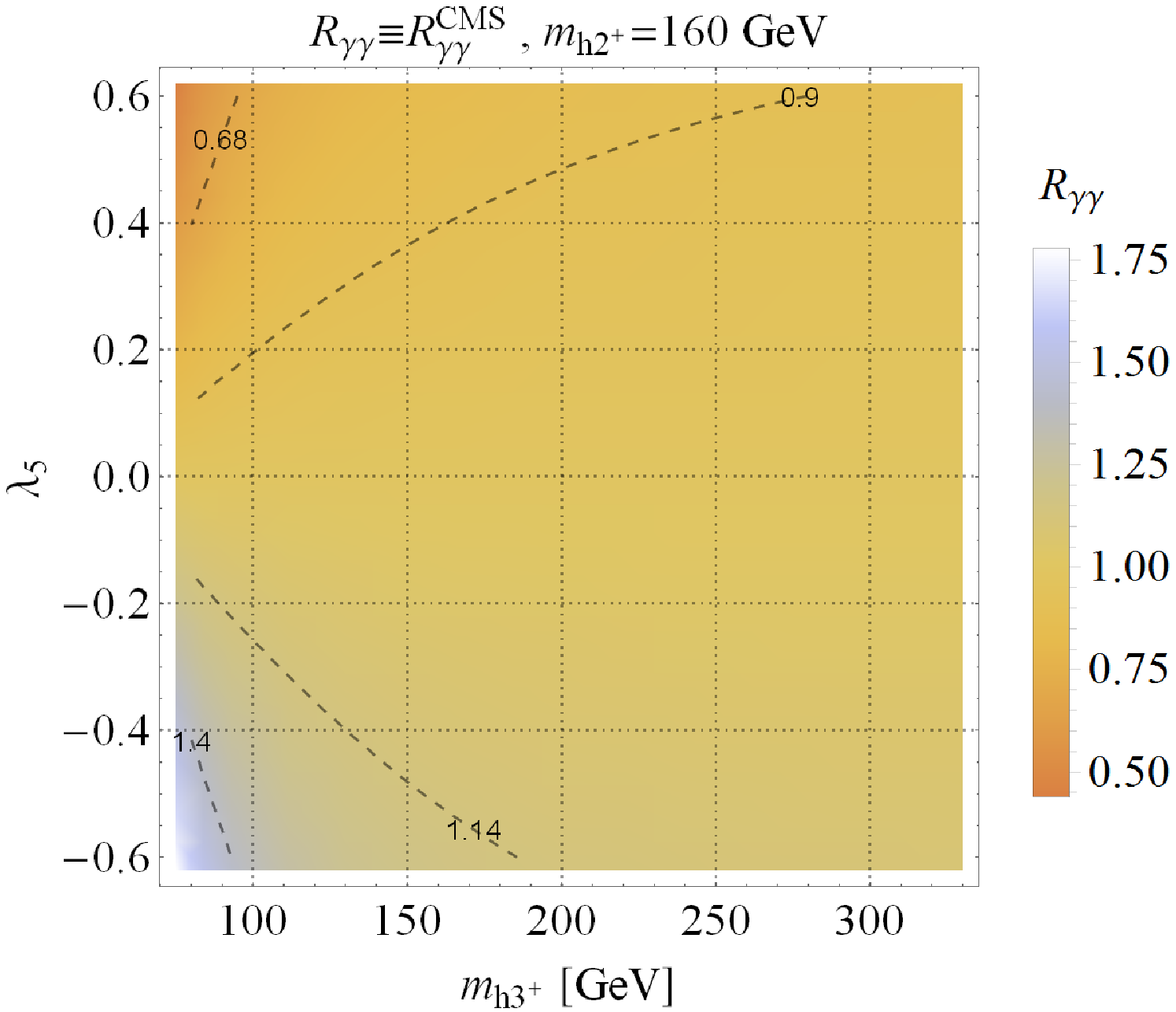}} \\
\subfloat[]{\includegraphics[width=8cm]{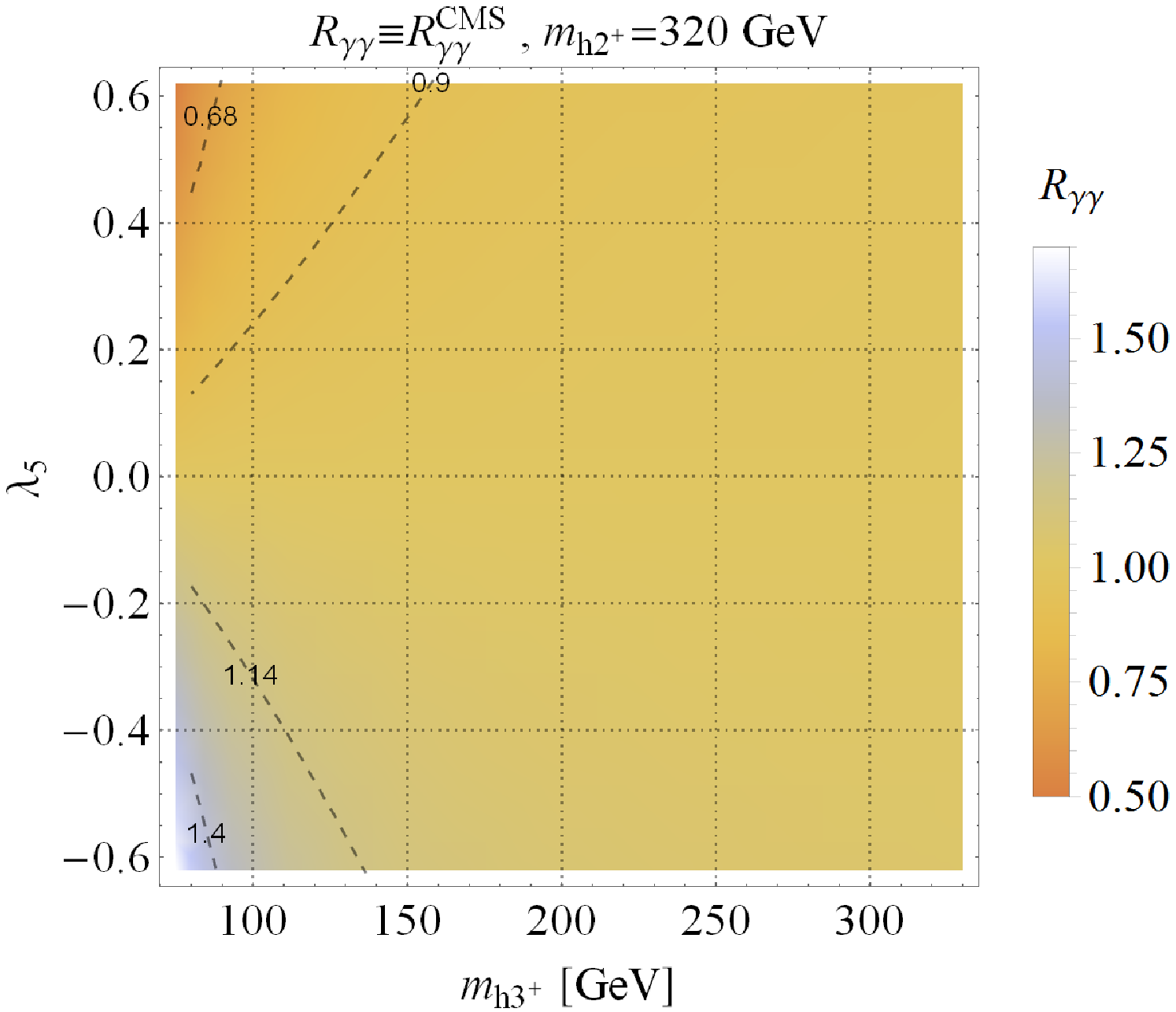}}
\caption{Specific values of $R_{\gamma\gamma}\equiv R_{\gamma\gamma}^\text{CMS}=1.14_{-0.23}^{+0.26}$ \cite{Khachatryan:2014ira} around the central value considering some $\sigma$ deviations, with the cases (a) $m_{h_2^+}=80$ GeV, (b) $m_{h_2^+}=160$ GeV, and (c) $m_{h_2^+}=320$ GeV, with $m_{h_3^+}\geq 80$ GeV and $-0.6\leq\lambda_5\leq 0.6$ .}\label{FIGURES-Results-Set-4}
\end{center}
\end{figure}

In order to test strictly our parameters we make a direct comparison of our $R_{\gamma\gamma}$ with the CMS data, namely,  $R_{\gamma\gamma}\big(\lambda_5,m_{h_{2,3}^+}\big)=R_{\gamma\gamma}^\text{CMS}=1.14_{-0.23}^{+0.26}$, for this we seek the values for which $\lambda_5$ and $m_{h_{2,3}^+}$ satisfy the experimental central value and also $\pm1\sigma$ deviations around it, where $+1\sigma=0.26$ and $-1\sigma=0.23$. When considering $-1\sigma$ then $R_{\gamma\gamma}$ downs to 0.91, and for $+1\sigma$ reaches 1.40.
In the Fig.~\ref{FIGURES-Results-Set-4}(a) with $m_{h_2^+}=80$ GeV, within $-0.6\leq\lambda_5\leq 0.6$ and $m_{h_3^+}\geq m_{h_2^+}$ the curves show the set of parameter values which meet the expectations for $R_{\gamma\gamma}^\text{CMS}$, we have also considered some sigma deviations of our interest for testing our parameters; in Fig.~\ref{FIGURES-Results-Set-4}(b) is presented the case $m_{h_2^+}=160$ GeV, and in (c) $m_{h_2^+}=320$ GeV.

Regarding to the channel $h\to\gamma Z$, now we can predict the $R_{\gamma Z}$ behaviour from the $R_{\gamma\gamma}$ graphs given in the Fig.~\ref{FIGURES-Results-Set-4} due to the dependence on common parameters.
For this we evaluate in $R_{\gamma Z}$ the set of values which trace the curves for $R_{\gamma\gamma}$ in the Fig.~\ref{FIGURES-Results-Set-4}, and in the Table~\ref{Sigma-Deviations-TABLE} we present the predictions for $R_{\gamma Z}$.
We have found that for $m_{h_{2,3}^+}\geq 80$ GeV and $-0.6\leq\lambda_5\leq 0.6$ occurs a constant correlation between both channels: respect to de central value with $m_{h_{2}^+}=80$ GeV our prediction is $R_{\gamma Z}=1.06$, and with $m_{h_2^+}\geq 160$ GeV is $R_{\gamma Z}=1.05$, and considering $-1\sigma$ the supression is 0.96 and for $+1\sigma$ rises to 1.16. In the Table~\ref{Sigma-Deviations-TABLE} we also report $m_{h_{2}^+}=$ 240 and 400 GeV, despite we do not plot them in the Fig.~\ref{FIGURES-Results-Set-4}, but we consider them important for presenting the constant correlated behavior.

\section{Conclusions}
\label{sec:con}

In this work we have considered the SM-like Higgs scalar decaying in $\gamma\gamma$ and $\gamma Z$ in the context of the IDM$S_3$ model which has also candidates for DM.
Both decays may have ratios $R_{\gamma\gamma}$ and $R_{\gamma Z}$ that can be enhanced or suppressed compared to the values predicted by the SM. The signal of the $\lambda_5$ parameter is the most responsible for this positive or negative deviation, Figs.~(\ref{FIGURES-Results-Set-1})-(\ref{FIGURES-Results-Set-4}). The shape and behavior of the curves of the both processes are very similar, and the difference of them is due to the massive particle in the final state of  $h\to \gamma Z$ channel. Therefore it is expected that the latter process be less sensitive than the two photons channel related to the common parameters $\lambda_5$ and $m_{h^+_{2,3}}$.
The lower value $m_{h^+_3}>25$ GeV was obtained from the limit established by the $Z\to h_3^+h_3^-$ invisible decay. Thus, our parameters are safe by considering this limit. We would like to stress that in the present model both charged scalars $h^+_{2,3}$ do not couple to fermions, they are inert, fact that highly simplifies the study of the impact of such new spin-0 content on the $h\to\gamma\gamma, \gamma Z$ processes. Due that they do not couple with fermions the lower limit obtained by LEP and LHC does not apply in this case. However, for at least one scalar, we use $m_{h^+_2}>80$ GeV from ATLAS~\cite{Aad:2013hla}.

Following the results from CMS \cite{Khachatryan:2014ira} for $R_{\gamma\gamma}$, we have explored the scenarios for the parameters $\lambda_5$, $m_{h^+_{2}}$ and $m_{h^+_{3}}$ which satisfies specific $R_{\gamma\gamma}$ values considering the experimental sigma deviation. We have concentrated our scenarios within $-0.6\leq\lambda_5\leq0.6$, $m_{h^+_{2}}\geq 80$ GeV and $m_{h^+_{3}}\geq 25$ GeV.
Worth to mention that $-0.4\leq\lambda_5\leq 0.4$ is consistent with our results of Ref.~\cite{Fortes:2014dca} where we had shown reasonable values of this model that can accommodate DM candidates.

Regarding the signal of the $\lambda_5$ parameter, we would like to call the attention to a similar analysis that had been done in the context of a general three Higgs doublets with $S_3$ symmetry, but without inert doublets, in Ref.~\cite{Das:2014fea}. In that case, both decays only have suppressions compared to the SM value: $R_{\gamma\gamma}\in [0.42, 0.80]$  and $R_{\gamma Z}\in [0.73, 0.93]$.
The difference between the analysis presented here and the one of Ref.~\cite{Das:2014fea} is that in our case the $\mu^2_d$
parameter does not contribute to the spontaneous symmetry breaking. In our analysis, the masses of the scalars are not limited by $v^2_{SM}$ and by  $\lambda$'s of the scalar potentials, this allow positive and negative values for $\lambda_5$, whereas in \cite{Das:2014fea} the respective parameter is always negative, see their Eq.~(46). Our analysis is congruent with theirs when our $\lambda_5$ is positive. An earlier analysis, also about the parameter space of both ratios in the IDM model, can be found in \cite{Krawczyk:2013pea}.

In our IDM$S_3$ a constant correlation occurs between the two processes when considering the scenario $m_{h_{2,3}^+}\geq 80$ GeV and $-0.6\leq\lambda_5\leq 0.6$ , this fact enable us to predict $R_{\gamma Z}$ from a given $R_{\gamma\gamma}$. Therefore, the comparison of our $R_{\gamma\gamma}$ with the $R_{\gamma\gamma}^\text{CMS}=1.14_{-0.23}^{+0.26}$ allow us to make such predictions, they are given in the Table~\ref{Sigma-Deviations-TABLE}: respect to de central value when considering $m_{h_{2}^+}=80$ GeV, our prediction is $R_{\gamma Z}=1.06$, and when $m_{h_2^+}\geq 160$ GeV is $R_{\gamma Z}=1.05$; besides, for $m_{h_2^+}\geq 80$ GeV if consider $+1\sigma$ the ratio reaches 1.16, while for $-1\sigma$ yields 0.96.
This kind of behavior has been observed in other multi-Higgs models which include  real \cite{Arina:2014xya} or complex \cite{Chen:2014lla} triplets, so this seems to be a general feature of multi-Higgs models.

Otherwise, the experimental reports on the $h\to\gamma Z$ decay will continue offering upper limits of one order of magnitude greater than the SM prediction, as commented in the Introduction, and it is expected  that at LHC Run 6 \cite{Dawson:2013bba,LHC-Runs-Talks} it reaches a luminosity of 3000 fb$^{-1}$ of $pp$ collisions and then could measure this mode with a precision of $54-57\%$ at ATLAS and of $20-24\%$ at CMS. What if an important increment is detected in the future reports? One possible answer to this question could be that maybe this is due to new physics effects that possible require a different coupling of the new particle with the $Z$ boson. For sure it will be an invitation to revisit the status of the SM.
In the SM, the decay $h\to\gamma Z$ is essentially due to the virtual $W$ gauge boson contribution, and the destructive interference caused by the top quark is not very significant, therefore the search of a deviation in this process is unlikely due to a possible correction in the $Zf\bar{f}$ vertices, besides the decay $Z\to f\bar{f}$ is well known.

\acknowledgments

ACBM thanks CAPES for financial support. ECFSF and JM thanks to FAPESP for financial support under the respective processes number 2011/21945-8 and 2013/09173-5. VP thanks to CNPq  for partial support. JM is grateful with Daniel Alva for useful discussions.

\appendix

\section{Form factors and the Passarino-Veltman scalar functions}
\label{appendix-form-factors}

Here we present explicitly the form factors $A_\text{Spin}$ \cite{Djouadi:2005gi,{Djouadi:2005gj}}, given in
Eqs.~(\ref{width-h-2gamma}) and (\ref{width-h-gammaZ}), in terms of the $B_0$ and $C_0$ Passarino-Veltman scalar functions \cite{'tHooft:1978xw}. We have constructed and solved each loop diagram with the Passarino-Veltman reduction method \cite{Passarino:1978jh} using \texttt{FeynCalc}~\cite{Mertig:1990an}, and also obtained the corresponding analytical solutions for the $B_0$ and $C_0$ scalar integrals via the Feynman parametrization method and dimensional regularization scheme \cite{Passarino:1978jh,Peskin:1995ev,Bardin:1999ak,Bohm:2001yx}. The solutions have been verified numerically
using \texttt{LoopTools} \cite{Hahn:1998yk}.
We have refrain from showing the construction of the loop integrals of the processes because they are frequently presented in the literature, instead we write down in detail the final result of the tensorial amplitudes, since they are usually omitted in terms of the Passarino-Veltman functions and even more unknown are their general analytical solutions which we found more practical for numerical evaluation, that is, without the need of splitting them in cases.

The one-loop decay $h\to\gamma\gamma$ is a low order process, therefore it is UV finite as there is no tree-level $h\gamma\gamma$ coupling in the lagrangian, since the SM is a renormalizable theory hence counterterms $h\gamma\gamma$ can not be present. Same argument applies to $h\to\gamma Z$ on the absence of $h\gamma Z$ interaction.

For the $h\to\gamma\gamma$ decay, with configuration $h(p_3)\to \gamma_{\mu_1}(p_1)\gamma_{\mu_2}(p_2)$,
the amplitude is
\begin{equation}\label{}
\mathcal{M}_{h\to\gamma\gamma} = \mathcal{M}_{\gamma\gamma}^{\mu_1\mu_2} \epsilon_{\mu_1}^*(\vec{p}_1,\lambda_1)\epsilon_{\mu_2}^*(\vec{p}_2,\lambda_2) \ ,
\end{equation}
with kinematics $p_3=p_1+p_2$, $p_3^2=m_h^2$, $p_1^2=p_2^2=0$, $p_1\cdot p_2=m_h^2/2$, and transversality conditions
$p_1\cdot\epsilon^*(\vec{p}_1,\lambda_1)=p_2\cdot\epsilon^*(\vec{p}_2,\lambda_2)=0$, this is, $p_1^{\mu_1}=p_2^{\mu_2}=0$.
The tensorial amplitude is
\begin{eqnarray}\label{}
\mathcal{M}_{\gamma\gamma}^{\mu_1\mu_2} &=& -i\frac{\sqrt{\sqrt{2}G_F}~\alpha}{2\pi}
    \left[ \sum_{i=1}^9 N_C^{f_i}Q_{f_i}^2 A_{1/2}^{\gamma\gamma}(\tau_{f_i}) +A_1^{\gamma\gamma}(\tau_W)
    +\frac{\lambda_5v_{SM}^2}{2}\sum_{i=2}^3\frac{1}{m_{h_i^+}^2}A_0^{\gamma\gamma}(\tau_{h_i^+})  \right]
    \nonumber\\
    && \times \left(\frac{m_h^2}{2} g^{\mu_1\mu_2}-p_2^{\mu_1}p_1^{\mu_2}\right) \ ,
\end{eqnarray}
where $\tau_X\equiv m_h^2/4m_X^2$ and $X= f, W, h^\pm$,
which satisfies electromagnetic gauge invariance via the accomplishment of the Ward identities  $p_{1\mu_1}\mathcal{M}_{\gamma\gamma}^{\mu_1\mu_2}=p_{2\mu_2}\mathcal{M}_{\gamma\gamma}^{\mu_1\mu_2}=0$.
The form factors are
\begin{eqnarray}\label{}
A_{1/2}^{\gamma\gamma} &\equiv& \frac{4m_f^2}{m_h^2}\left[ 2+ \left( 4m_f^2-m_h^2 \right)C_0^{h,f} \right] \nonumber\\
    &=& 2[\tau_f+(\tau_f-1)f(\tau_f)]\tau_f^{-2} \ ,
\end{eqnarray}
\begin{eqnarray}\label{}
A_1^{\gamma\gamma} &\equiv& -2\left\{1+6\frac{m_W^2}{m_h^2}\left[ 1+\left( 2m_W^2-m_h^2\right)C_0^{h,W}\right] \right\} \nonumber\\
    &=& -[2\tau_W^2+3\tau_W+3(2\tau_W-1)f(\tau_W)]\tau_W^{-2} \ ,
\end{eqnarray}
\begin{eqnarray}\label{}
A_0^{\gamma\gamma} &\equiv& -\frac{4m_{h^+}^2}{m_h^2}\left(1+2m_{h^+}^2C_0^{h,h^+} \right) \nonumber\\
    &=& -[\tau_{h^+}-f(\tau_{h^+})]\tau_{h^+}^{-2} \ ,
\end{eqnarray}
\begin{equation}\label{function-f-Djouadi}
f(\tau) \equiv \left\{
    \begin{array}{lcl}
      \arcsin^2\sqrt{\tau} & , & \tau\leq 1 \\
    -\frac{1}{4}\left(\log \frac{1+\sqrt{1-\tau^{-1}}}{1-\sqrt{1-\tau^{-1}}}-i\pi\right)^2 & , &   \tau > 1
    \end{array}
    \right. \ , \ \tau\equiv \frac{m_h^2}{4 m_X^2} \ .
\end{equation}
The three-point Passarino-Veltman scalar function is
\begin{eqnarray}\label{PaVe-C0-case1}
C_0^{h,X} &\equiv& C_0(0,0,m_h^2,m_X^2,m_X^2,m_X^2) \nonumber\\
    &=& \frac{1}{2m_h^2} \log^2 \left(-\frac{1+\sqrt{1-\frac{4(m_X^2-i\epsilon)}{m_h^2}}}{1-\sqrt{1-\frac{4(m_X^2-i\epsilon)}{m_h^2}}} \right) \nonumber\\
    &=& -\frac{2}{m_h^2}\arctan^2\frac{-i}{\sqrt{1-\frac{4(m_X^2-i\epsilon)}{m_h^2}}} \nonumber\\
    &=& -\frac{2}{m_h^2}f\left(\tau \right) \ .
\end{eqnarray}

For the $h\to\gamma Z$ decay, with configuration $h(p_3)\to \gamma_{\mu_1}(p_1) Z_{\mu_2}(p_2)$, the amplitude is
\begin{equation}\label{}
\mathcal{M}_{\gamma Z} = \mathcal{M}_{\gamma Z}^{\mu_1\mu_2} \epsilon_{\mu_1}^*(\vec{p}_1,\lambda_1)\epsilon_{\mu_2}^*(\vec{p}_2,\lambda_2) \ ,
\end{equation}
with kinematics $p_3=p_1+p_2$, $p_3^2=m_h^2$, $p_1^2=0, \ p_2^2=m_Z^2$, $p_1\cdot p_2=(m_h^2-m_Z^2)/2$, and transversality conditions $p_1\cdot\epsilon^*(\vec{p}_1,\lambda_1)=p_2\cdot\epsilon^*(\vec{p}_2,\lambda_2)=0$, this is, $p_1^{\mu_1}=p_2^{\mu_2}=0$.
The tensorial amplitude is
\begin{eqnarray}\label{}
\mathcal{M}_{h\to\gamma Z}^{\mu_1\mu_2} &=& i\frac{\sqrt{\alpha} G_F m_W}{\sqrt{2}\pi^{3/2}}
    \left[ \frac{2}{c_W} \sum_{i=1}^9 N_C^{f_i}Q_{f_i}g_V^{f_i} A_{1/2}^{\gamma Z}(\tau_{f_i}) +A_1^{\gamma Z}(\tau_W)
    +\frac{\lambda_5v_{SM}^2v_{h^\pm}}{2}\sum_{i=2}^3\frac{1}{m_{h_i^+}^2}A_0^{\gamma Z}(\tau_{h_i^+}) \right] \ ,
    \nonumber\\
    && \times \left( \frac{m_h^2-m_Z^2}{2} g^{\mu_1\mu_2}-p_2^{\mu_1}p_1^{\mu_2} \right) \ ,
\end{eqnarray}
where $\tau_X\equiv 4m_X^2/m_h^2$, $\lambda_X\equiv 4m_X^2/m_Z^2$ and $X= f, W, h^\pm$, which satisfies $U(1)_\text{em}$ gauge invariance through the fulfilment of the Ward identity for the photon $p_{1\mu_1}\mathcal{M}_{\gamma Z}^{\mu_1\mu_2}=0$. Moreover, in this process is also  satisfied the Ward identity for the $Z$ boson $p_{2\mu_2}\mathcal{M}_{\gamma Z}^{\mu_1\mu_2}=0$.
The form factors are
\begin{eqnarray}\label{}
A_{1/2}^{\gamma Z} &=& -\frac{2m_f^2}{m_h^2-m_Z^2}-\frac{2m_Z^2m_f^2}{(m_h^2-m_Z^2)^2}\left(B_0^{h,f}-B_0^{Z,f}\right)
    +m_f^2\left(1-\frac{4m_f^2}{m_h^2-m_Z^2} \right) C_0^{h,Z,f} \nonumber\\
    &=& I_1(\tau_f,\lambda_f)-I_2(\tau_f,\lambda_f) \ ,
\end{eqnarray}
\begin{eqnarray}\label{}
A_1^{\gamma Z} &=& \frac{1}{m_Wm_Z(m_h^2-m_Z^2)^2} \left\{
    [2m_W^2(m_h^2+6m_W^2)-m_Z^2(m_h^2+2m_W^2)] \left[m_h^2-m_Z^2+m_Z^2\left(B_0^{h,W}-B_0^{Z,W}\right)\right] \right. \nonumber\\
    && \left. +2(m_h^2-m_Z^2)m_W^2[6m_W^2(2m_W^2+m_Z^2)-2m_Z^4+m_h^2(m_Z^2-6m_W^2)] C_0^{h,Z,W} \right\} \nonumber\\
    &=& c_W \left\{4(3-t_W^2)I_2(\tau_W,\lambda_W)+[(1+2\tau_W^{-1})t_W^2-(5+2\tau_W^{-1})]I_1(\tau_W,\lambda_W) \right\} \ ,
\end{eqnarray}
\begin{eqnarray}\label{}
A_0^{\gamma Z} &=& \frac{2m_{h^+}^2}{(m_h^2-m_Z^2)^2} \left[m_Z^2\left(B_0^{h,h^+}-B_0^{Z,h^+}\right)
    +(m_h^2-m_Z^2) \left(1+2m_{h^+}^2C_0^{h,Z,h^+}\right) \right] \nonumber\\
    &=&-I_1(\tau_{h^+},\lambda_{h^+}) \ ,
\end{eqnarray}
and the auxiliary functions
\begin{eqnarray}\label{auxiliar-functions}
I_1(\tau,\lambda) &\equiv& \frac{\tau\lambda}{2(\tau-\lambda)}+\frac{\tau^2\lambda^2}{2(\tau-\lambda)^2}[f(\tau^{-1})-f(\lambda^{-1})]
    +\frac{\tau^2\lambda}{(\tau-\lambda)^2}[g(\tau^{-1})-g(\lambda^{-1})] \ , \nonumber\\
I_2(\tau,\lambda) &\equiv&  -\frac{\tau\lambda}{2(\tau-\lambda)}[f(\tau^{-1})-f(\lambda^{-1})] \ .
\end{eqnarray}
We emphasize that here in $h\to\gamma Z$ is used $\tau\equiv 4m_X^2/m_h^2$, opposite to the $h\to\gamma\gamma$ case, therefore here $f(\tau^{-1})$ evaluates $\tau^{-1}=m_h^2/4m_X^2$ for consistency because originally $f(\tau)$ is defined in Eq.~(\ref{function-f-Djouadi}) with $\tau \equiv m_h^2/4m_X^2$, and exactly the same situation holds for
\begin{equation}\label{function-g-Djouadi}
g(\tau) \equiv \left\{
    \begin{array}{lcl}
    \sqrt{\tau^{-1}-1}\arcsin\sqrt{\tau} & , & \tau\leq 1  \\
    \frac{\sqrt{1-\tau^{-1}}}{2}\left(\log \frac{1+\sqrt{1-\tau^{-1}}}{1-\sqrt{1-\tau^{-1}}}-i\pi\right) & , & \tau > 1
    \end{array}
    \right. \ , \ \tau\equiv \frac{m_h^2}{4m_X^2} \ ,
\end{equation}
where we disagree with the inequalities orientations given in Refs.~\cite{Djouadi:2005gi,{Djouadi:2005gj}}, but the correct definition can be found in the same author's Ref.~\cite{Spira:1995rr}, as also used in Ref.~\cite{Swiezewska:2012eh}.

The two-point scalar function, with its ultraviolet (UV) divergent term $\Delta$, is
\begin{eqnarray}
B_0^{h,X} &=& B_0(m_h^2,m_X^2,m_X^2) \nonumber\\
    &=& \Delta-\log\frac{m_X^2}{\mu^2}+2-\sqrt{1-\frac{4(m_X^2-i\epsilon)}{m_h^2}}
    \log \left(-\frac{1+\sqrt{1-\frac{4(m_X^2-i\epsilon)}{m_h^2}}}{1-\sqrt{1-\frac{4(m_X^2-i\epsilon)}{m_h^2}}} \right)                  \nonumber\\
    &=& \Delta-\log\frac{m_X^2}{\mu^2}+2-2\ i\sqrt{1-\frac{4(m_X^2-i\epsilon)}{m_h^2}}
    \arctan\frac{-i}{\sqrt{1-\frac{4(m_X^2-i\epsilon)}{m_h^2}}} \nonumber\\
    &=& \Delta-\log\frac{m_X^2}{\mu^2}+2-2~g(\tau) \ ,
\end{eqnarray}
\begin{equation}\label{divergence}
\Delta \equiv \frac{2}{4-D}-\gamma_E+\log 4\pi \ ,
\end{equation}
where $g(\tau)$ is accordingly with Eq.~(\ref{function-g-Djouadi}), and the difference of two $B_0$ with same virtual masses yields the UV-finite result
\begin{equation}\label{B0-finite}
B_0^{h,X}-B_0^{Z,X}=-2[g(\tau)-g(\lambda)] \ .
\end{equation}
The \texttt{LoopTools} program evaluates any $B_0$ without the $\Delta+\log\mu^2$ term by default because in any UV-finite process such terms must vanish, e.g. Eq.~(\ref{B0-finite}).

Finally, the last three-point scalar function is
\begin{eqnarray}\label{PaVe-C0-case2}
C_0^{h,Z,X} &\equiv& C_0(0,m_h^2,m_Z^2,m_X^2,m_X^2,m_X^2) \nonumber\\
    &=&\frac{m_h^2C_0^{h,X}-m_Z^2C_0^{Z,X}}{m_h^2-m_Z^2}\nonumber\\
    &=& \frac{-2}{m_h^2-m_Z^2} [f(\tau)-f(\lambda)] \ ,
\end{eqnarray}
where $C_0^{h,X}$ and $C_0^{Z,X}$ are given in Eq.~(\ref{PaVe-C0-case1}).


\newpage

\begin{table}[ht]
\begin{tabular}{l|c|c}
\hline\hline
Ratio               & ATLAS   & CMS   \\ \hline
$R_{\gamma\gamma}$  & 1.17$_{-0.27}^{+0.27}$  \cite{Aad:2014eha}  & 1.14$_{-0.23}^{+0.26}$ \cite{Khachatryan:2014ira} \\
$R_{\gamma Z}$      & $<11\quad\;\;$         \cite{Aad:2014fia}   & $ < 9.5\quad\;\,$  \cite{Chatrchyan:2013vaa}\\
\hline\hline
\end{tabular}
\caption{Ratios of the experimental measured values compared to the SM predictions reported by ATLAS and CMS. In this work we use the CMS data for the two photons process because it gives the more stringent deviation.}\label{Atlas-CMS-Rates-TABLE}
\end{table}


\newpage


\newpage


\begin{table}[!h]
  \centering
\begin{tabular}{|c|c|c|c|c|c|c|}\hline
\multirow{2}{*}{Deviation} & \multirow{2}{*}{$R_{\gamma\gamma}$} & \multicolumn{5}{c|}{$R_{\gamma Z}$}  \\
\cline{3-7}
                        &      & $m_{h_2^+}=80$ GeV & $m_{h_2^+}=160$ GeV & $m_{h_2^+}=240$ GeV & $m_{h_2^+}=320$ GeV & $m_{h_2^+}=400$ GeV \\
\hline
$+2\sigma$            & 1.66 & 1.26 & -    & -    & -    & -    \\
$+1\sigma$            & 1.40 & 1.16 & 1.16 & 1.16 & 1.16 & 1.16 \\
$\quad 0 \ \sigma$      & 1.14 & 1.06 & 1.05 & 1.05 & 1.05 & 1.05 \\
$-1\sigma$            & 0.91 & 0.96 & 0.96 & 0.96 & 0.96 & 0.96 \\
$-2\sigma$            & 0.68 & 0.86 & 0.86 & 0.86 & 0.86 & 0.86 \\
\hline
\end{tabular}
\caption{
Predictions for $R_{\gamma Z}$ from $R_{\gamma\gamma}\equiv R_{\gamma\gamma}^{\text{CMS}}=1.14_{-0.23}^{+0.26}$ \cite{Khachatryan:2014ira} around the central value considering $\sigma$ deviations, within $-0.6\leq\lambda_5\leq 0.6$, $m_{h_3^+}\geq 80$ GeV and with different fixed values from $m_{h_2^+}\geq 80$ GeV. The evolution of $\lambda_5$ and $m_{h_3^+}$ can be seen explicitly in the Fig.~\ref{FIGURES-Results-Set-4} for the cases $m_{h_2^+}=$ 80, 160, and 320 GeV.
The number absence means prediction out of the $\lambda_5$ interval.}\label{Sigma-Deviations-TABLE}
\end{table}

\end{document}